\documentclass[sigconf]{acmart}

\makeatletter
\def\@ACM@checkaffil{% Only warnings
	\if@ACM@instpresent\else
	\ClassWarningNoLine{\@classname}{No institution present for an affiliation}%
	\fi
	\if@ACM@citypresent\else
	\ClassWarningNoLine{\@classname}{No city present for an affiliation}%
	\fi
	\if@ACM@countrypresent\else
	\ClassWarningNoLine{\@classname}{No country present for an affiliation}%
	\fi
}
\makeatother

\AtBeginDocument{%
	}

\copyrightyear{2023}
\acmYear{2023}
\setcopyright{acmlicensed}\acmConference[CVMP '23]{European Conference on Visual Media Production}{November 30-December 1, 2023}{London, United Kingdom}
\acmBooktitle{European Conference on Visual Media Production (CVMP '23), November 30-December 1, 2023, London, United Kingdom}
\acmPrice{15.00}
\acmDOI{10.1145/3626495.3626503}
\acmISBN{979-8-4007-0426-0/23/11}

\usepackage{booktabs} % For formal tables

\usepackage{amsfonts,amssymb}
\usepackage{multirow}

\begin{document}
	
	%%
	%% The "title" command has an optional parameter,
	%% allowing the author to define a "short title" to be used in page headers.
	\title[ITM-LUT: AI Optimized LUT for Inverse Tone-mapping]{Redistributing the Precision and Content in 3D-LUT-based Inverse Tone-mapping for HDR/WCG Display}
	
	%%
	%% The "author" command and its associated commands are used to define
	%% the authors and their affiliations.
	%% Of note is the shared affiliation of the first two authors, and the
	%% "authornote" and "authornotemark" commands
	%% used to denote shared contribution to the research.
	
	\author{Cheng Guo}
	\affiliation{%
		\institution{State Key Laboratory of Media Convergence and Communication, Communication University of China}
		%\country{China}
	}
	%\streetaddress{No.2 Xingke 1st. St.}
	%\city{Shenzhen}
	\email{guocheng@cuc.edu.cn}
	
	\author{Leidong Fan}
	\affiliation{%
		\institution{School of Electronic and Computer Engineering, Shenzhen Graduate School, Peking University}
		%\country{China}
	}
	%\streetaddress{No.2 Xingke 1st. St.}
	%\city{Shenzhen}
	\email{fanleidong@stu.pku.edu.cn}
	
	\author{Qian Zhang}
	\affiliation{%
		\institution{Academy of Broadcasting Planning}
		\city{Beijing}
		\country{China}
	}
	%\streetaddress{No.2 Xingke 1st. St.}
	%\city{Shenzhen}
	\email{zhangfanqian@139.com}
	
	\author{Hanyuan Liu}
	\affiliation{%
		\institution{Academy of Broadcasting Planning}
		\city{Beijing}
		\country{China}
	}
	%\streetaddress{No.2 Xingke 1st. St.}
	%\city{Shenzhen}
	\email{lhy2871@126.com}
	
	\author{Kanglin Liu}
	\affiliation{%
		\institution{Peng Cheng Laboratory}
		\city{Shenzhen}
		\country{China}
	}
	%\streetaddress{No.2 Xingke 1st. St.}
	%\city{Shenzhen}
	%\country{China}
	\email{max.liu.426@gmail.com}
	
	\author{Xiuhua Jiang}
	\affiliation{%
		\institution{Peng Cheng Laboratory}
		\city{Shenzhen}
		\country{China}
	}
	%\streetaddress{No.2 Xingke 1st. St.}
	\email{jiangxiuhua@cuc.edu.cn}

	%%
	%% By default, the full list of authors will be used in the page
	%% headers. Often, this list is too long, and will overlap
	%% other information printed in the page headers. This command allows
	%% the author to define a more concise list
	%% of authors' names for this purpose.
	\renewcommand{\shortauthors}{Guo et al.}
	
	%%
	%% The abstract is a short summary of the work to be presented in the
	%% article.
	\begin{abstract}
		ITM (inverse tone-mapping) converts SDR (standard dynamic range) footage to HDR/WCG (high dynamic range /wide color gamut) for media production.
		It happens not only when remastering legacy SDR footage in front-end content provider, but also adapting on-the-air SDR service on user-end HDR display.
		The latter requires more efficiency, thus the pre-calculated LUT (look-up table) has become a popular solution.
		Yet, conventional fixed LUT lacks adaptability, so we learn from research community and combine it with AI.
		Meanwhile, higher-bit-depth HDR/WCG requires larger LUT than SDR, so we consult traditional ITM for an efficiency-performance trade-off:
		We use 3 smaller LUTs, each has a non-uniform packing (\textit{precision}) respectively denser in dark, middle and bright luma range.
		In this case, their results will have less error only in their own range, so we use a contribution map to combine their best parts to final result.
		With the guidance of this map, the elements (\textit{content}) of 3 LUTs will also be redistributed during training.
		We conduct ablation studies to verify method's effectiveness, and subjective and objective experiments to show its practicability.
		Code is available at: \url{https://github.com/AndreGuo/ITMLUT}.
	\end{abstract}
	
	%%
	%% The code below is generated by the tool at http://dl.acm.org/ccs.cfm.
	%% Please copy and paste the code instead of the example below.
	%%
	
	
	%%
	%% Keywords. The author(s) should pick words that accurately describe
	%% the work being presented. Separate the keywords with commas.
	%% A "teaser" image appears between the author and affiliation
	%% information and the body of the document, and typically spans the
	%% page.
	\begin{teaserfigure}
		\includegraphics[width=\textwidth]{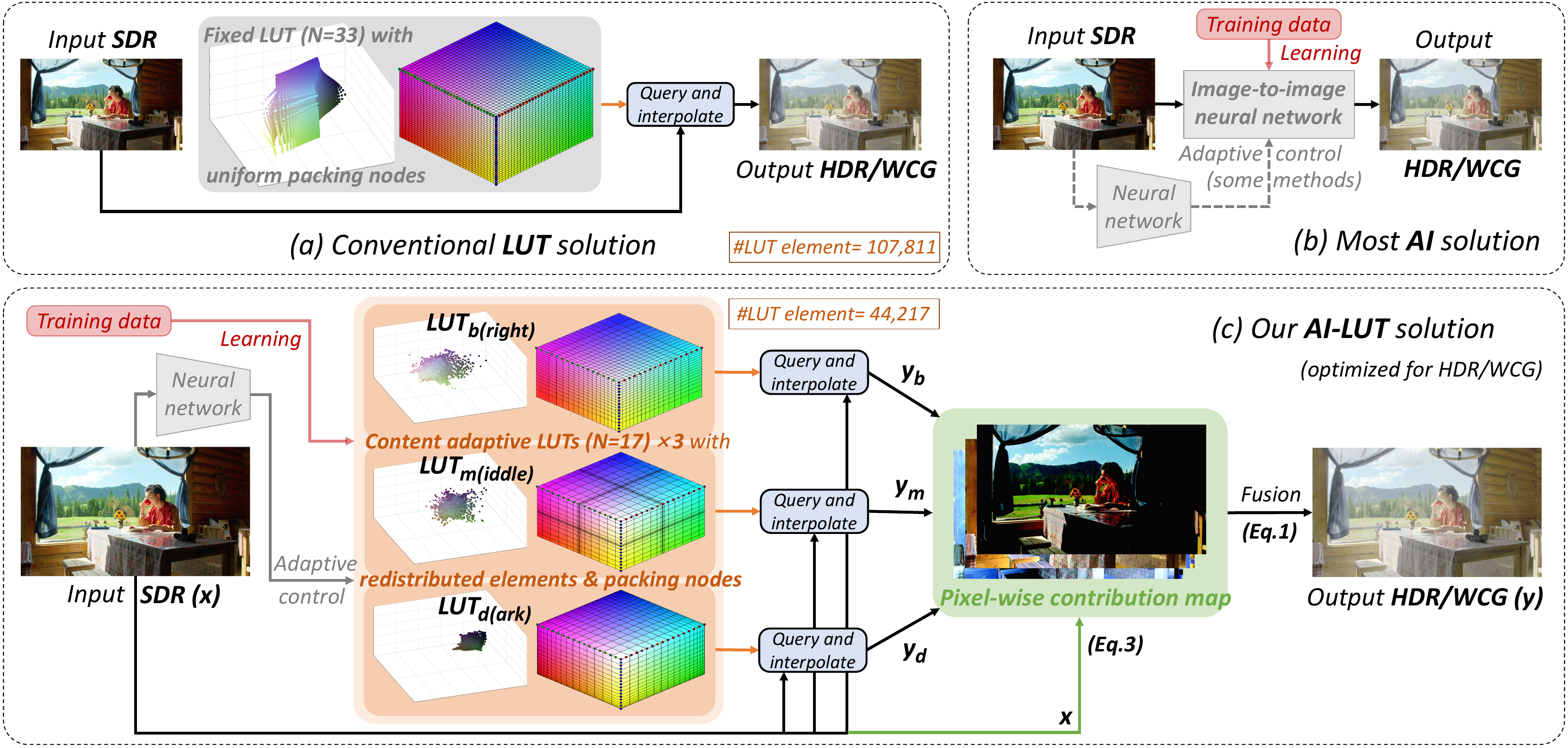}
		\caption{Different ITM (inverse tone-mapping) solutions.
			Conventional LUT (a) lacks adaptability. While for user-end application, end-to-end AI methods (b) fail for efficiency.
			Therefore, we combine AI with LUT: our LUT is learned form training data, and varies with input ($\mathbf{x}$).
			Also, we use 3 smaller LUTs (N=17) with explicitly-defined non-uniform packing precision (colored cube) respectively denser in dark, middle and bright range.
			Then, a pixel-wise contribution map is used to combine the best part of 3 results ($\mathbf{y}_b$, $\mathbf{y}_m$ and $\mathbf{y}_d$), and the output content of 3 LUTs (colored dots) will also be redistributed during learning.
			In this case, we reach similar performance to single bigger LUT (N=33), yet occupying fewer \#param of LUT (44217<107811).}
		\Description{Teaser Figure.}
		\label{fig:teaser}
	\end{teaserfigure}
	
	%\received{20 February 2007}
	%\received[revised]{12 March 2009}
	%\received[accepted]{5 June 2009}
	
	%%
	%% This command processes the author and affiliation and title
	%% information and builds the first part of the formatted document.
	\maketitle
	
	% ===============================================================
	\section{Introduction}
	In media industry, ultra-high definition (UHD) surpasses high-definition (HD) in 5 dimensions: resolution, frame-rate, bit-depth, dynamic range and color gamut.
	The first 3 belong to \textit{definition} which gives a more precise description, while the last 2 bring advances in \textit{expressiveness} via larger luminance and color container:
	Image/video with >1000\textit{nit} max-luminance in PQ/HLG non-linearity and BT.2020 gamut is defined as high dynamic range/wide color gamut (HDR/WCG), while that in 100\textit{nit} max-luminance and BT.709 gamut is termed standard dynamic range (SDR).
	
	With the proliferation of HDR/WCG display in consumer market, conflicts lies in 2 aspects:
	Content providers are eager to start HDR/WCG service while such content is still scarce.
	Also, users are expecting to enjoy better experience via readily-available HDR/WCG display while most service is still in SDR.
	This entails inverse tone-mapping (ITM) which converts SDR to HDR/WCG.
	% \cite{Kim19,Zeng20,Chen211,Xu222,Guo23} \textit{etc}, 
	Most AI-ITMs (\S\ref{sec:ai_itm}) aim at remastering legacy SDR footage, they mainly focus on recovering degradation by end-to-end network, and fail for user-end-display scenario since an edge device (\textit{e.g.} set-to-box) is with limited computation resources.
	Thus, a 3D look-up table (LUT) suits precisely, since it pre-calculates all $\mathbb{R}^3 \rightarrow \mathbb{R}^3$ mapping and replaces runtime computation with caching, indexing and interpolation.
	
	Yet, conventional LUT requires designer's expertise in \textit{e.g.} color science for a \textit{top-down} design and couldn't embrace the merits of big-data in a \textit{bottom-up} learning.
	Also, fixed LUT lacks adaptability and is hard to follow cinephotographer's diverse technical/artistic intent between SDR and HDR/WCG in real workflow.
	Hence, we combine LUT with AI:
	We train the method with latest refined HDR/WCG dataset\cite{Guo23}, and endow it with adaptability by merging 5 basic LUTs\cite{20-3DLUT} using network-generated weights.
	
	While most AI-LUTs (\S\ref{sec:ai_lut}) are oriented for 8-bit SDR, 10/12-bit HDR/WCG requires larger LUT size, \textit{i.e.} less efficiency, for the same error level\cite{3D-LUT-on-HDR}.
	Therefore, we must utilize LUT's \textit{precision} of sparse representation (packing vertices) more efficiently.
	One solution is assigning non-uniform packing within single LUT. For example, \cite{LUT-optim-HDR-WCG} has denser packing at lower range based on that HDR/WCG's numerical distribution concentrates more on smaller codewords than SDR, while \cite{22-AdaInt} lets packing vertices 
	gather at the range where AI think is with more non-linearity.
	However, for method's generalization, we can't assume the numerical distribution of SDR and expect the method to interpolate less error only on such SDR.
	
	To this end, we consult traditional ITM \textit{e.g.} \cite{PITM}, and use 3 LUTs with distinct contribution on dark, middle and bright luma ranges.
	For each LUT, its non-uniform packing on SDR's RGB cube (\textit{precision}) is assigned denser vertices on specific range, by Eq.\ref{eq:vertices}, so its interpolation result will have less error only on corresponding range.
	In this case, we use pixel-wise map (Eq.\ref{eq:lum_prob}) to let each result contribute their best range to final result.
	Also, with the constraint of this map, LUT's output on HDR/WCG's RGB cube (\textit{content}) will also concentrate on their own responsible ranges.
	
	By redistributing the \textit{precision} and \textit{content} of 3 LUTs (N=17), we have more efficient utilization than conventional (AI-)LUT (N=33), and reach an acceptable result compared to other AI solutions.
	Experiments are as follows:
	The first is ablation study on packing vertices (\textit{precision} redistribution), contribution map (guidance of \textit{content} redistribution) and LUT initialization \textit{etc.}
	The rest are comparative experiments using both HDR/WCG-tailored metrics and subjective experiment  \textit{etc.}
	Our contributions are:
	
	\begin{itemize}
		\item In the filed of AI (learning-based) ITM, to the best of our knowledge, our method is the first 3D-LUT-based, and currently the fastest under UHD/4K resolution.
		\item In AI (learning-based) LUT, we make the first attempt adapting higher-bit-depth HDR/WCG, and discuss the impact of some rarely-studied ingredients \textit{e.g.} LUT initialization.
	\end{itemize}
	
	% ===============================================================
	\section{Related Work}
	\label{sec:related_work}
	
	% ---------------------------------------------------------------
	\subsection{Traditional Inverse Tone-mapping}
	\label{sec:trad_itm}
	
	\cite{AkyuzEO} use fixed $\gamma$ curve to expand SDR's luminance to HDR, while \cite{MasiaEO,BistITM} add parameter(s) controlled by SDR statistics, and \cite{LuzardoITM} further improve the shape of $\gamma$ curve.
	Some algorithms still apply global expansion, yet use piecewise function based on different perceptual characteristics of bright and dark luminance ranges: \cite{MeylanEO} use polygonal function, \textit{Method C} in \cite{BT2446} set 58.5\textit{nit} knee-point and use a 2-piece linear-exponential curve, \cite{ChenITM} use 2-piece 2-order polynomial with real-time pivot search, while \cite{PITM} apply 3-piece 1-order polynomial expansion directly in PQ non-linear domain.
	
	Rather a numerical different expansion, some methods use spatial extra operator.
	\cite{BanterleEO} expand SDR by inverting existing tone-mapping operator\cite{Reinhard02TMO} and combining with SDR residual by a weight map, SDR highlight area is then expanded by a guidance map from density estimation.   
	\cite{RempelEO} use simpler expansion curve, yet more complicated guidance map by image pyramid, this guidance map is later improved by \cite{KOEO}.
	\cite{HuoEO} use Retina-theory-assisted curve affected by not only SDR's global statistics, but also neighboring pixels.
	
	Yet, traditional ITM algorithms:
	(1) only define an $\mathbb{R}^1$$\rightarrow$$\mathbb{R}^1$ mapping on luminance, thus chrominance components need to be borrowed/adjusted from SDR to HDR.
	(2) are incapable recovering degradation.
	(3) have user-controlled parameters \textit{i.e.} lacked adaptability.
	(4) will not expand color volume when comes to HDR/WCG, and no advanced WCG pixels will be produced if only color space conversion (CST)\cite{BT2087} is appended, rather gamut expansion. % \textit{e.g.} \cite{AndersonGE,TakeuchiGE,GamutNet}.
	
	% ---------------------------------------------------------------
	\subsection{Learning-based (AI) Inverse Tone-mapping}
	\label{sec:ai_itm}
	
	Based on above deficiencies, researchers have turned to AI approach which conduct $\mathbb{R}^3$$\rightarrow$$\mathbb{R}^3$ mapping directly from SDR to HDR/WCG.
	Some AI method \textit{e.g.}\cite{Eilertsen17SIHDR} is for image-based lighting, not media industry, so we follow the taxonomy in \cite{CheatHDR,Guo23} to call them single-image HDR reconstruction (SI-HDR).
	Here, a method is termed ITM if its output is claimed in PQ/HLG and BT.2020 gamut.
	
	First, global operations still play an important role in AI:
	\cite{Mustafa22,Kim201} are pixel-independent, while \cite{Chen211,Shao22,Xu23,Chen231,Kim23,Chen232} use multi-step network where at least one sub-step belongs to global operation.
	Meanwhile, \cite{Mustafa22,Yao23} use reversible operation.
	Modulation is popular obtaining adaptability in AI-ITM: modulation in \cite{Chen211,Xu222,Xu23} is broadcast among all pixels, while that of \cite{He221,He222,Shao22,Xu223,Chen232} is spatially different.
	For video application, \cite{Xu192,Cao22,Zou20,Xu221,Xu23,Xu223} improve the temporal coherency processing consecutive frames, \cite{Kim18,Kim19,Kim202,Zeng20,Xu221,He222,Yao23,Zhang23} conduct joint ITM and super-resolution considering HD/UHD resolution discrepancy.
	\cite{Kim201,Chen212,Tang22} utilize traditional ITM as bypass or pre-processing to reduce the workload of AI part.
	Other motivations include:
	\cite{Xu191} is for badly-exposed SDR, and \cite{Cheng22ITMDM,Guo23} explore the impact of degradation model in training \textit{etc}.
	
	We further conclude their intended \textbf{application scenario} in Tab.\ref{tab:itm}.
	As seen, the majority is designed for front-end remastering.
	
	\begin{table*}[h]
		\setlength{\abovecaptionskip}{0cm}
		\setlength{\belowcaptionskip}{0cm}
		\caption{Motivation and application scenario of AI-ITM: most are for front-end remastering, not our user-end display.}
		\centering
		\scriptsize
		\begin{tabular}{|c|c|c|c|}
			\hline
			\textbf{Claimed application scenario} & \textbf{Front-end remastering}      & \textbf{Undeclared} & \textbf{User-end display (ours)}  \\ \hline
			Multi-step \& global operation          & \cite{Chen211,Shao22,Xu23,Chen231,Chen232}  & -  & \cite{Mustafa22,Kim201,Kim23}  \\ \hline
			Using reversible operation   & -    & -   & \cite{Mustafa22,Yao23} \\ \hline
			Using modulation    & \cite{Chen211,Xu222,Xu23,Shao22,Xu223,Chen232} & \cite{He221}        & \cite{He222} \\ \hline
			Improving temporal coherency    & \cite{Xu192,Zou20,Xu223,Xu23}                       & \cite{Cao22,Xu221}  & -                      \\ \hline
			Joint super-resolution    & \cite{Kim18,Kim19,Kim202,Zeng20,Zhang23}         & \cite{Xu221}   & \cite{He222,Yao23}     \\ \hline
			
			Combining traditional method          & \cite{Chen212,Tang22}                    & -                   & \cite{Kim201}          \\ \hline
			
			Other motivation  & \cite{Xu191,Guo23,Cheng22ITMDM}   & -                   & -   \\ \hline
		\end{tabular}
		\label{tab:itm}
	\end{table*}

	\begin{table*}[t]
		\setlength{\abovecaptionskip}{0cm}
		\setlength{\belowcaptionskip}{0cm}
		\caption{Status-quo of AI-3D-LUT. We streamline them to \textbf{4 ingredients}:
			(1) how the \textbf{expressiveness} of learned LUT(s) is improved, by the number of basic LUTs (\textbf{\#basicLUT}, \textit{col.}2) and introducing extra dimension (\textbf{extraD}, \textit{col.}4) with certain number of possible values (\textbf{\#}) \textit{etc.},
			(2) what the \textbf{neural network} is designed to learn (\textit{col.}5),
			(3) \textbf{packing} strategy (\textit{col.}6) and
			(4) \textbf{interpolation} (\textit{col.}7).
		}
		\centering
		\small
		\begin{tabular}{|c|c|c|c|c|c|c|} 
			\hline
			\multirow{2}{*}{\textbf{AI-LUT methods}} & \multicolumn{3}{c|}{\textbf{Expressiveness of the trained LUT(s)}} & \multirow{2}{*}{\begin{tabular}[c]{@{}c@{}}\textbf{Output of}\\\textbf{neural network(s)}\end{tabular}} & \multirow{2}{*}{\textbf{\begin{tabular}[c]{@{}c@{}}\textbf{Packing intervals}\\\textbf{(input sampling)}\end{tabular}}} & \multirow{2}{*}{\begin{tabular}[c]{@{}c@{}}\textbf{Inter-}\\\textbf{polation}\end{tabular}}  \\ 
			\cline{2-4}
			& \textbf{\#basicLUT} & \textbf{LUTsize} & \textbf{extraD (\#)}    &           &               &                                 \\ 
			\hline
			\textbf{A3D-LUT}\scriptsize{\cite{20-3DLUT}}        & 3$\times$1        & 3$\times$33$^3$               & -               & weights (of basicLUTs)     & uniform                  & trilinear                       \\ 
			\hline
			\textbf{SA-LUT-Nets}\scriptsize{\cite{21-SA-3DLUT}}   & 3$\times$10       & 3$\times$33$^3$               & `category' (10)    & weights \& category map    & uniform                  & trilinear                       \\ 
			\hline
			\textbf{4D-LUT}\scriptsize{\cite{22-4D-LUT}}         & 3$\times$1        & 3$\times$33$^4$               & `context' (33)     & weights \& context map     & uniform                  & quadrilinear                    \\ 
			\hline
			\textbf{CLUT-Net}\scriptsize{\cite{22-CLUT-Net}}       & 20$\times$1       & 3$\times$5$\times$20          & -               & weights                    & uniform                  & trilinear                       \\ 
			\hline
			\textbf{NLUT}\scriptsize{\cite{23-NLUT}}              & 2048$\times$1     & 3$\times$32$\times$32          & -               & weights                    & uniform                  & trilinear                       \\ 
			\hline
			\textbf{F2D-LUTs}\scriptsize{\cite{22-F2D-LUT}}       & 6$\times$3        & 2$\times$33$^2$               & chan. order (3)     & weights                    & uniform                  & bicubic                         \\
			\hline
			\textbf{DualBLN}\scriptsize{\cite{22-DualBLN}}        & 5$\times$1        & 3$\times$36$^3$               & -               & LUTs fusion map            & uniform                  & trilinear                       \\ 
			\hline
			\textbf{AdaInt}\scriptsize{\cite{22-AdaInt}}         & 3$\times$1        & 3$\times$33$^3$               & -               & weights \& intervals       & learned non-uniform       & trilinear                       \\ 
			\hline
			\textbf{SepLUT}\scriptsize{\cite{22-SepLUT}}         & no                & 3$\times$9/17$^3$    & -        & directly 1D \& 3D LUT    & learned pre-nonlinear    & trilinear                       \\ 
			\hline
			\textbf{ITM-LUT} (ours)        & 5$\times$3        & 3$\times$17$^3$               & luma prob. (3)    & weights       & explicitly-defined non-uniform       & trilinear                     \\
			\hline
		\end{tabular}
		\label{tab:3d_luts}
	\end{table*}
	
	% ---------------------------------------------------------------
	\subsection{Learning-based (AI) 3D-LUT}
	\label{sec:ai_lut}
	
	Since above remastering-oriented end-to-end AI methods are too bulky for user-end, we turn to global (pixel-independent) AI operation with better efficiency:
	\cite{Glob-Yan16} predicts pixel's polynomial basis function, \cite{Glob-Gharbi17} learns per-pixel affine transformation coefficient in bilateral space, \cite{Glob-Guo20,Glob-Bianco20} estimate channel's global curve independently (no cross-channel contamination).
	\cite{GlobRL-Hu18,GlobRL-Park18} predict the decision on an all-global image processing pipeline (ISP), and \cite{Glob-He20,Glob-Wang22} use $1\times1$ convolution (no receptive filed).
	Considering the expressiveness and efficiency of each approach, we finally turn to AI-3D-LUT.
	
	So, we first investigate the status-quo of AI-LUT:
	Except non-3D LUTs for super-resolution\cite{21-SR-LUT,22-MuLUT,23-Online}, medical imaging\cite{21-SA-LuT-Nets} and multi-exposure fusion\cite{23-MEFLUT} \textit{etc.}, the rest (Tab.\ref{tab:3d_luts}) are all 3D LUTs of $RGB$ triplet which enable color manipulation.
	We only describe their ideas and refer to \S9 of \cite{Kang06Book} for LUT basics.
	The first AI 3D LUT\cite{20-3DLUT} merges 3 basic LUTs to an adaptive one using neural-network-generated weights (basic LUTs are also trainable). 
	Since LUT consists of output's sparse expression (\textit{content}) and its input packing (\textit{precision}), subsequent work lies in 2 aspects:
	
	The first technical route is the expressiveness of LUT \textit{content}: \cite{21-SA-3DLUT} merges 3$\times$10 basic LUTs to 10 and predict 10-channel pixel-wise category map to guide the weighted sum of 10 look-up result.
	\cite{22-4D-LUT} is similar and use network-predicted `context' (another dimension) map to merge 3 basic 4D LUTs conducting $\mathbb{R}^4$$\rightarrow$$\mathbb{R}^3$ mapping.
	\cite{22-DualBLN} flattens 3D LUTs to 2D images before adaptive multi-layer fusion.
	Meanwhile, LUT's efficiency is also ameliorated: \cite{22-CLUT-Net} finds a compressed representation of 3-D LUT, so they can merge up to 20 basic LUTs with minimum overhead.
	With this compressed representation, \cite{23-NLUT} increases the number of basic LUTs to 2048 for stronger adaptability needed in photorealistic style transfer.
	While \cite{22-F2D-LUT} decouples a 3D LUT into several 2D LUTs (respectively on R-G, R-B and G-B channel) with fewer size.
	
	The second idea is on LUT's packing \textit{precision}:
	\cite{22-SepLUT} applies 3 learned 1D LUTs before 3D look-up, which resembles \S9.4 of \cite{Kang06Book} and equals to 3D LUT with non-uniform packing interval.
	\cite{22-AdaInt} shares similar idea, but different source of non-uniformity: they directly set each vertices interval controllable, similar to \cite{LUT-node-optim,LUT-optim-HDR-WCG}.

	% ===============================================================
	\section{Proposed Method}
	\label{sec:method}

	% ---------------------------------------------------------------
	\subsection{Module Design}
	\label{sec:model}
	
	Given $\mathbf{x} \in \mathbb{R}^{3\times h\times w}$ the single frame of input SDR in \textit{gamma} non-linearity and BT.709 \textit{RGB} primaries (gamut),  $\mathbf{y} \in \mathbb{R}^{3\times h\times w}$ the output HDR/WCG frame in PQ non-linearity and BT.2020 gamut, our method (Fig.\ref{fig:teaser}(c)) can be expressed as below:
	\begin{equation}
		\label{eq:overall}
		\mathbf{y}=\sum_{\tiny{l\in\{b,m,d\}}}^{}{p_l(\mathbf{x})\cdot \mathbf{y}_l,\ \ \  \mathbf{y}_l=interp.(\mathbf{x},LUT_l,\mathbf{v}_l)}
	\end{equation}
	where $l\in\{b,m,d\}$ represents specific LUT on \textbf{b}right, \textbf{m}iddle or \textbf{d}ark \textbf{l}uma range,
	$p_l(\mathbf{x}) \in \mathbb{R}^{3\times h\times w}$ is the per-pixel \& per-channel contribution map for the weighted fusion of 3 look-up results $\mathbf{y}_l$ ($\mathbf{y}_b$, $\mathbf{y}_m$ and $\mathbf{y}_d$) $\in \mathbb{R}^{3\times h\times w}$, determined by `luma probability'.
	
	\subsubsection{Precision redistribution}
\begin{figure}[h]
		\centering
		\includegraphics[width=0.67\linewidth]{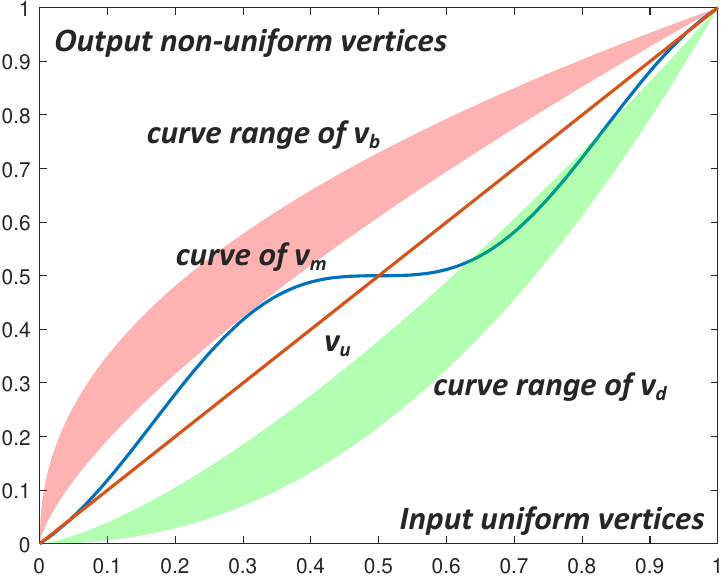}
		\caption{The non-linear curves (Eq.\ref{eq:vertices}) to redistribute uniform vertices (horizontal axis) to non-uniform (vertical axis).}
		\label{fig:vertices}
	\end{figure}

	For each range $l$, $interp.(\cdot_l)$ stands for look-up and trilinear interpolation on $\mathbf{x}$ using the learned $LUT_l$ with packing \textbf{v}ertices $\mathbf{v}_l$.
	Keeping the number of vertices along [0,1] (\textit{i.e.} LUT size $N$) unchanged, \cite{LUT-node-optim,LUT-optim-HDR-WCG,22-AdaInt} tell us that conventional vertices can be set to non-uniform, to reach denser representation and consequently less interpolation error in specific range.
	Therefore, one of our discriminate treatment of 3 luma ranges is that each LUT's $\mathbf{v}_l$ is designed to have different non-uniformity:
	\begin{equation}
		\label{eq:vertices}
		\begin{array}{l} 
			\left\{\begin{array}{l} 
				\mathbf{v}_b={\mathbf{v}_u}^{1/(1.4+0.8\overline{\mathbf{x}})} \\ 
				\mathbf{v}_d={\mathbf{v}_u}^{2.2-0.8\overline{\mathbf{x}}} \\ 
				\mathbf{v}_m={\dfrac{3\pi\mathbf{v}_u-cos(3\pi\mathbf{v}_u)+1}{3\pi+2}} 
			\end{array}\right. \ ,\mathbf{v}_u=linspace(0,1,N)
		\end{array} 
	\end{equation}
	where $\overline{\mathbf{x}}$ is $\mathbf{x}$'s arithmetic mean on specific R/G/B channel, it endows $\mathbf{v}_b$ and $\mathbf{v}_d$ with adaptability.
	As seen, we transfer uniform vertices $\mathbf{v}_u \in \mathbb{R}^{N\times 1}$ to non-uniform $\mathbf{v}_l$ by curves in Eq.\ref{eq:vertices} and Fig.\ref{fig:vertices}.
	
	As in Fig.\ref{fig:vertices}, these curves will make $\mathbf{v}_b$/$\mathbf{v}_m$/$\mathbf{v}_d$ respectively denser near $\mathbf{x}$=1/0.5/0.
	Meanwhile, since SDR($\mathbf{x}$)-HDR/WCG($\mathbf{y}$) relation is monotonically-increasing, such $\mathbf{v}_b$/$\mathbf{v}_m$/$\mathbf{v}_d$ will also reach more precise result respectively in $\mathbf{y}$'s higher/middle/lower range.
	
	\subsubsection{Contribution redistribution}
	However, specific $\mathbf{v}_l$ will only boost performance in its targeted range, \textit{e.g.} $\mathbf{v}_d$ has denser representation near $\mathbf{x}$'s 0 and consequently less error on $\mathbf{y}$'s lower range, yet sparser near $\mathbf{x}$'s 1 and worse on $\mathbf{y}$'s rest higher range.
	Our solution is making the final result $\mathbf{y}$ a combination of $\mathbf{y}_b$/$\mathbf{y}_m$/$\mathbf{y}_d$'s own better ranges: As in Eq.\ref{eq:overall}, we use $\mathbf{x}$'s `luma probability' $p_l \in \mathbb{R}^{3\times h\times w}$ (Fig.\ref{fig:abl_lum_prob}) to determine the per-pixel contribution of $\mathbf{y}_b$, $\mathbf{y}_m$ and $\mathbf{y}_d$:
	\begin{equation}
		\label{eq:lum_prob}
		\begin{array}{l} 
			\left\{\begin{array}{l}
				p_b(\mathbf{x})=clamp(\frac{\mathbf{x}-t_b}{1-t_b},0,1) \\ 
				p_d(\mathbf{x})=clamp(\frac{\mathbf{x}-t_d}{0-t_d},0,1) \\ 
				p_m(\mathbf{x}) = 1-p_b(\mathbf{x})-p_d(\mathbf{x}) 
			\end{array}\right. \ ,\mathbf{x}\in[0,1]
		\end{array} 
	\end{equation}
	where threshold $t_b$=0.55 \& $t_d$=0.45 controlling each $\mathbf{y}_l$'s contribution are determined in \S\ref{sec:abl}(2).
	For example, $LUT_d$ ($\mathbf{y}_d$) will not contribute to the higher $\mathbf{y}$ whose contribution should belong to $LUT_b$.
	Note that $p_b$+$p_m$+$p_d$=1 $\forall \ \mathbf{x}\in[0,1]$, so weighted result $p_l(\mathbf{x})\cdot \mathbf{y}_l$ can be directly added ($\sum$ in Eq.\ref{eq:overall}) without scaling.
	In this case, \textit{e.g.} pixel $\mathbf{x}_{i}$=0.3 is with $[p_b,p_m,p_d]$=[0, 1/3, 2/3], so corresponding $\mathbf{y}_{i}$ comes in 0, 1/3, 2/3 numerical portion from $\mathbf{y}_b$, $\mathbf{y}_m$ and $\mathbf{y}_d$.
	
	\subsubsection{Content Redistribution}
	Till now, we have redistributed the \textit{precision} of 3 N=17 LUTs distinctly to bright, middle and dark ranges (Eq.\ref{eq:vertices}), and the contribution redistribution (Eq.\ref{eq:lum_prob}) serves as a consequent compensation of their own precision neglect in the rest range.
	With AI, their \textit{content} (vertices' corresponding output $\mathbf{y}$, Fig.\ref{fig:teaser} colored dots) will also be redistributed during training.
	
	As seen, each $LUT_l$ concentrates on own range.
	The reason is that under the guidance of contribution map (Eq.\ref{eq:lum_prob}), \textit{e.g.} $\mathbf{y}_d$ will contribute nothing when $\mathbf{x}_{i}$>0.45, so any output value corresponding to $\mathbf{x}_{i}$>0.45 in $LUT_d$ will become invalid, thus $LUT_d$ will be trained to abandon them and concentrate on valid dark range.
	
	\subsubsection{LUT adaptability.}
	From above we know that any final $LUT_l$ can have a redistributed \textit{content} with the constraint of Eq.\ref{eq:lum_prob} and training, the current issue is how final $LUT_l$ should be generated.
	One option is \cite{22-SepLUT} who set LUT's 3$\times$N$^3$ elements as trainable parameters.
	The drawback is that once LUT is trained, it becomes fixed and will not alter with $\mathbf{x}$'s statistics.
	Hence most AI-ITM (Tab.\ref{tab:itm}) involve $\mathbf{x}$'s statistics by condition\cite{Glob-He20}, while \cite{Mustafa22} is by own latent space.
	
\begin{figure}[h]
		\centering
		\includegraphics[width=0.6\linewidth]{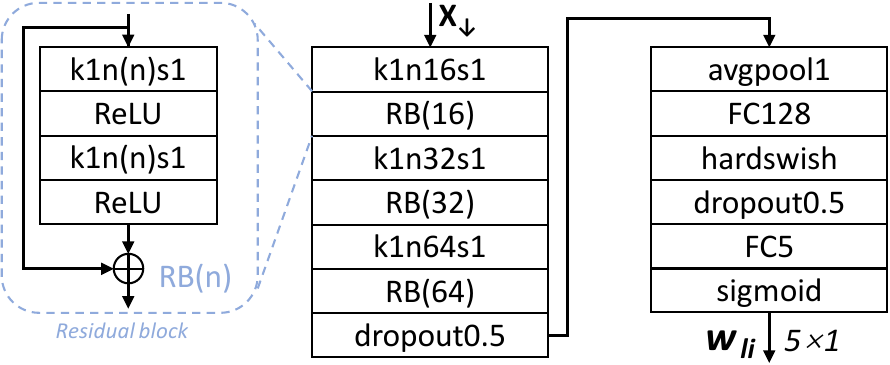}
		\caption{Structure of weight predictor ($NetW_l$):
			it consists of lightweight convolution and FC (fully-connected) layers, and is feed with 1/8 down-sampled input $\mathbf{x}_{\downarrow}$.
			Each $NetW_l$ ($NetW_b$, $NetW_m$ and $NetW_d$) share the same structure, not parameters.}
		\label{fig:network}
	\end{figure}

	Mechanisms above will not help LUT.
	Early attempt of LUT's adaptability is mapping SDR to 1D-LUT using regression\cite{KaduITM}, and for 3D-LUT, we finally consult \cite{20-3DLUT}:
	Each of $LUT_l$ is merged from 5 basic LUTs ($LUT_{li},i\in\{0,1,2,3,4\}$) using the neural-network-generated weights $([w_{l0},\ldots,w_{li}]^{\mathrm{T}})$:
	\begin{equation}
		\label{eq:lut_fuse}
		LUT_l=\sum_{i=0}^{4}{w_{li}LUT_{li}},\ [w_{l0},\ldots,w_{li}]^{\mathrm{T}}=NetW_l(\mathbf{x}_{\downarrow})
	\end{equation}
	In this case, network (Fig.\ref{fig:network}) can output varying weights for different $\mathbf{x}$, and consequently the $LUT_l$ will become different.
	Meanwhile, \textit{content} of 5 basic LUTs are also optimized during training.
	
	% ---------------------------------------------------------------
	\subsection{Training Strategy}
	\label{sec:training}
	
	\subsubsection{Initialization}
	All weights in $NetW_l$ are Xavier\cite{Xavier} initialized, rest parameters are Kaiming\cite{Kaiming} initialized.
	Note that 3$\times$5 basic LUTs $[LUT_{l0},\ldots,LUT_{l4}]$ are also trainable, most SDR-oriented AI-LUT (\S\ref{sec:ai_lut}) initialize them with $\mathbf{y}$=$\mathbf{x}$ and/or $\mathbf{y}$=1.
	We show in \S\ref{sec:abl}.(3) that this applies to retouching task whose $\mathbf{x}$ and $\mathbf{y}$ are with small numerical discrepancy, yet not ITM whose pixel value alters dramatically due to SDR-HDR/WCG container discrepancy.
	Therefore, as in Tab.\ref{tab:init_lut}, we initialize $LUT_{li}$ with 5 different LUTs which have already been applied in ITM:
	
	\begin{table}[h]
		\setlength{\abovecaptionskip}{0cm}
		\setlength{\belowcaptionskip}{0cm}
		\caption{Basic LUTs $LUT_{li}$ are mainly initialized with practical LUTs for ITM (visualized in Fig.\ref{fig:lut_init}).
			They consider HDR/WCG discrepancy, and cover different and reasonably large output ranges which will reduce deep leaning's search space.}
		\centering
		\small
		\begin{tabular}{|c|cc|}
			\hline
			\textbf{basicLUT} & \multicolumn{2}{c|}{\textbf{initialized with}}                                  \\ \hline
			\begin{tabular}[c]{@{}c@{}}$LUT_{l0}$\\ \tiny{$l \in \{b,m,d\}$}\end{tabular} & \multicolumn{1}{c|}{\textit{\textbf{C\_100DW}}} & \scriptsize{\begin{tabular}[c]{@{}c@{}}\textbf{c}ontainer conversion from \textit{gamma(1/0.45)} SDR to BT.2020\\ PQ HDR/WCG, w. SDR's \textbf{d}efuse \textbf{w}hite mapped to \textbf{100}\textit{nit},\\ the conversion in most software \textit{e.g.} \textit{macOS Compressor}.\end{tabular}} \\ \hline
			$LUT_{l1}$ & \multicolumn{1}{c|}{\textit{\textbf{OCIO2}}} & \scriptsize{\begin{tabular}[c]{@{}c@{}}`\textit{sRGB}' to `\textit{Rec.2020 ST2084 (1000nits)}' conversion provided\\ by \textbf{O}pen\textbf{C}olor\textbf{IO} v\textbf{2}.0\cite{OCIO2}.\end{tabular}} \\ \hline
			$LUT_{l2}$ & \multicolumn{1}{c|}{\textit{\textbf{C\_203DW}}} & \scriptsize{\begin{tabular}[c]{@{}c@{}}\textbf{c}ontainer conversion from \textit{gamma(1/0.45)} SDR to BT.2020\\ PQ HDR/WCG, w. SDR's \textbf{d}efuse \textbf{w}hite mapped to \textbf{203}\textit{nit}.\end{tabular}} \\ \hline
			$LUT_{l3}$ & \multicolumn{1}{c|}{\textit{\textbf{DaVinci}}} & \scriptsize{\begin{tabular}[c]{@{}c@{}}`\textit{Rec.709}/\textit{sRGB}' to `\textit{Rec.2020}/\textit{ST2084}' conversion provided\\ by grading software \textit{DaVinci}'s `\textit{color space conversion}'.\end{tabular}} \\ \hline
			$LUT_{l4}$ & \multicolumn{1}{c|}{\textit{\textbf{identity}}} & \scriptsize{LUT recording $\mathbf{y}=\mathbf{x}$ (identity mapping).} \\ \hline
		\end{tabular}
		\label{tab:init_lut}
	\end{table}
	
	\subsubsection{Loss function}
	Given $\mathbf{x}_{p}\in\mathbb{R}^{b\times3\times600\times600}$ ($b$=4 is batch size) the [0.25x,1.25x] random resized and randomly cropped \textbf{p}atch from training set's $\mathbf{x}$,  $\mathbf{y}_{p}$ the method output and  $\bar{\mathbf{y}}_{p}$ the GT-HDR counterpart of $\mathbf{x}_{p}$, our loss function can be formulated as:
	\begin{equation}
		\label{eq:loss}
		loss=\Vert\mathbf{y}_{p}-\bar{\mathbf{y}}_{p}\Vert_1+{\textstyle \sum_{l}}reg_{lut}(LUT_l)
	\end{equation}
	where $reg_{lut}=0.01reg_{smooth.}+10reg_{mono.}$ is 3D-LUT specialized smoothness and monotonicity terms from \cite{20-3DLUT}.
	
	\subsubsection{Training set}
	Training set has recently received more attention in ITM\cite{Cheng22ITMDM}, we use 3848 HDR/WCG frames from latest-refined HDRTV4K\cite{Guo23} dataset as label.
	However, SDR there contains more degradation simulating legacy SDR, while user-end-received SDR is generally with quality control \textit{i.e.} less degradation.
	Also, global operations including LUT are incapable recovering degradation.
	Therefore, we `degrade' HDR/WCG to SDR by down-conversion in \textit{DaVinci} software (rather HDRTV4K's original degradation models) so our input SDR contains less degradation thus $l_1$ loss will not compensate for unable-to-recover patterns.
	
	\begin{figure}[h]
		\centering
		\includegraphics[width=\linewidth]{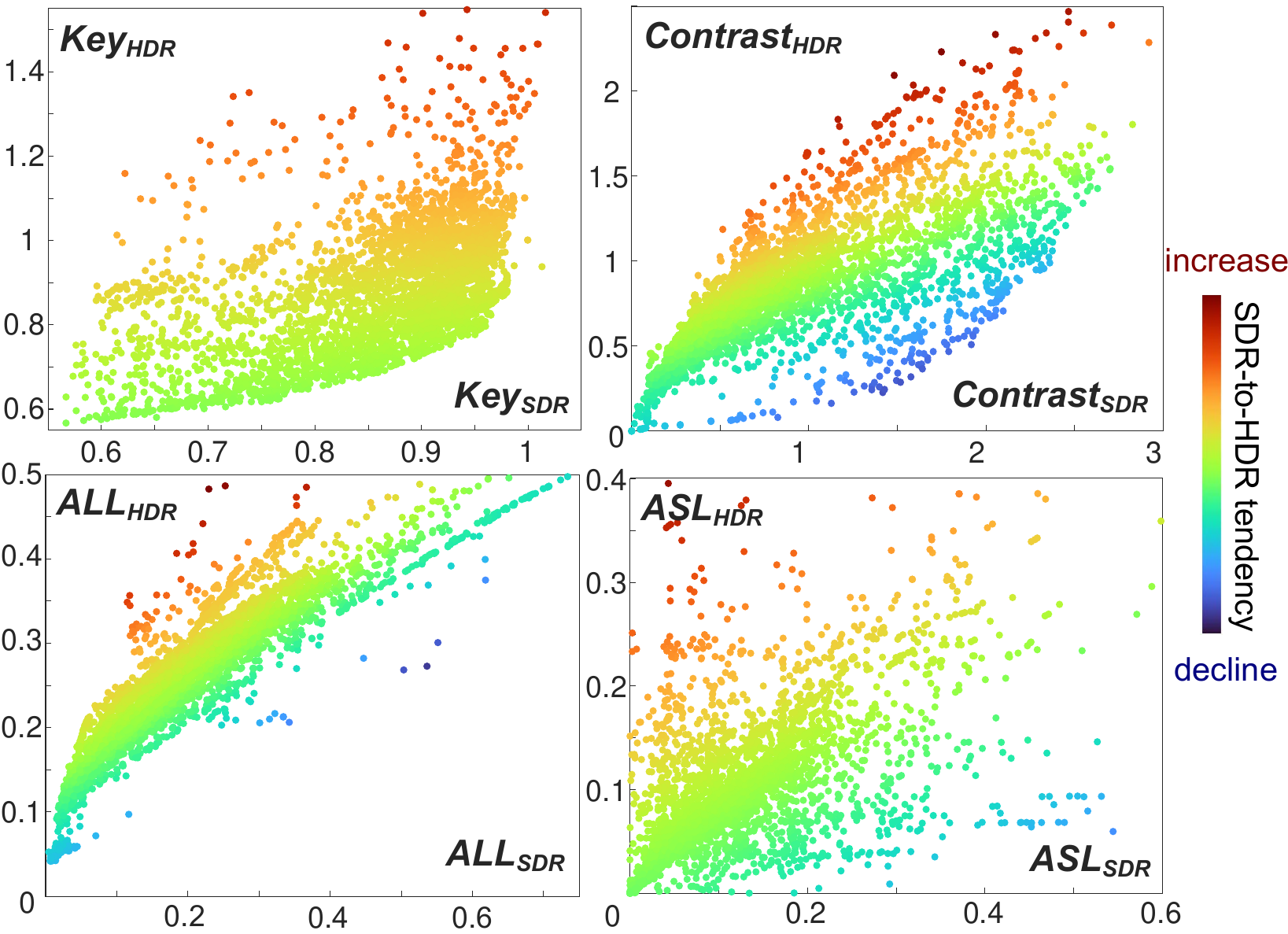}
		\caption{Training set's diverse SDR-HDR/WCG relationship, each dot is an SDR-HDR/WCG pair: \textit{e.g.} if SDR-HDR/WCG relation is fixed, dots should be around a single curve, and method will consequently learn a undesired fixed transform.}
		\label{fig:stat}
	\end{figure}
	
	Also, the prerequisite for AI to learn adaptability is that it already exists in training set: SDR-HDR/WCG relation should be diverse so method can learn a varying mapping.
	To this end, when using \textit{DaVinci} down-conversion, we ask the colorist to use different setting for different HDR/WCG, and this diverse relation is manifested by \textit{key value}\cite{LuzardoITM}, \textit{contrast}\cite{LuzardoITM}, \textit{ALL} (avg. luminance level) and \textit{ASL} (avg. saturation level) of different SDR-HDR/WCG pairs in Fig.\ref{fig:stat}.

	\subsubsection{Others}
	Each time dataloader is called, UHD SDR-HDR/WCG pairs are randomly resized and cropped to $600\times600$ patches.
	Method is based on \textit{PyTorch}, and optimized using AdaM and a decaying learning rate starting from 2$\times$10$^{-4}$.
	Training ends when epoch=35.
	
	% ===============================================================
	\section{Experiments}
	\label{sec:exp}
	
	% ---------------------------------------------------------------
	\subsection{Ablation Studies}
	\label{sec:abl}
	
	\begin{table*}[h]
		\setlength{\abovecaptionskip}{0cm}
		\setlength{\belowcaptionskip}{0cm}
		\caption{Configurations and quantitative results of ablation studies. $^{\star}$: Neural network is same as \cite{22-AdaInt}.}
		\centering
		\small
		\begin{tabular}{|cc|c|c|c|r|r|r|r|}
			\hline
			\multicolumn{2}{|c|}{\textbf{Ablations terms}}                                                                                                                            & \textbf{Packing vertices}         & \textbf{Luma Probability}          & \textbf{LUT initialization}                                                            & \multicolumn{1}{c|}{\textbf{PSNR}(dB)} & \multicolumn{1}{c|}{\textbf{SSIM}} & \multicolumn{1}{c|}{\textbf{$\Delta$E}} & \multicolumn{1}{c|}{\textbf{VDP3}} \\ \hline
			\multicolumn{1}{|c|}{original}   & \textbf{A}                                                                                                                         & non-uniform, Eq.\ref{eq:vertices}                  & \multirow{3}{*}{Eq.\ref{eq:lum_prob} ($t_b$=0.55, $t_d$=0.45)} & \multirow{8}{*}{\begin{tabular}[c]{@{}c@{}}5 different LUTs\\ (Tab.\ref{tab:init_lut})\end{tabular}}  & 34.203                                 & 0.9593                             & 17.013                           & 9.0363                             \\ \cline{1-3} \cline{6-9} 
			\multicolumn{1}{|c|}{\multirow{2}{*}{\begin{tabular}[c]{@{}c@{}}\textbf{\S\ref{abl1}}: non-uni-\\ form vertices\end{tabular}}}                                      & \textbf{1B} & by neural network$^{\star}$                 &                                         &                                                                                                          & 33.883                                 & 0.9576                             & 17.316                           & 8.9477                             \\ \cline{2-3} \cline{6-9} 
			\multicolumn{1}{|c|}{}                                                                                                                         & \textbf{1C} & uniform                           &                                         &                                                                                                          & 34.085                                 & 0.9586                             & 17.017                           & 9.0360                             \\ \cline{1-4} \cline{6-9} 
			\multicolumn{1}{|c|}{\multirow{5}{*}{\begin{tabular}[c]{@{}c@{}}\textbf{\S\ref{abl2}}: luma\\ probability map\\ (contribution\\ of 3 $\mathbf{y}_l$)\end{tabular}}} & \textbf{2B} & \multirow{7}{*}{non-uniform, Eq.\ref{eq:vertices}} & Eq.\ref{eq:lum_prob} ($t_b$=0.7, $t_d$=0.3)                    &                                                                                                          & 33.959                                 & 0.9575                             & 17.346                           & 9.0015                             \\ \cline{2-2} \cline{4-4} \cline{6-9} 
			\multicolumn{1}{|c|}{}                                                                                                                         & \textbf{2C} &                                   & Eq.\ref{eq:lum_prob} ($t_b$=0.9, $t_d$=0.1)                    &                                                                                                          & 33.986                                 & 0.9561                             & 17.572                           & 8.9571                             \\ \cline{2-2} \cline{4-4} \cline{6-9} 
			\multicolumn{1}{|c|}{}                                                                                                                         & \textbf{2D} &                                   & $p_b$=$p_m$=$p_d$=1/3                            &                                                                                                          & 32.429                                 & 0.9516                             & 20.574                           & 8.9435                             \\ \cline{2-2} \cline{4-4} \cline{6-9} 
			\multicolumn{1}{|c|}{}                                                                                                                         & \textbf{2E} &                                   & further-soft seg. (Eq.\ref{eq:lum_prob_soft})                                     &                                                                                                          & 33.424                                 & 0.9530                             & 18.105                           & 9.0585                             \\ \cline{2-2} \cline{4-4} \cline{6-9} 
			\multicolumn{1}{|c|}{}                                                                                                                         & \textbf{2F} &                                   & hard seg. ($t_b$=2/3, $t_d$=1/3)    &                                                                                                          & 30.938                                 & 0.9325                             & 24.200                           & 8.4941                             \\ \cline{1-2} \cline{4-5} \cline{6-9} 
			\multicolumn{1}{|c|}{\multirow{2}{*}{\begin{tabular}[c]{@{}c@{}}\textbf{\S\ref{abl3}}: LUT\\ initialization\end{tabular}}}                                          & \textbf{3B} &                                   & \multirow{2}{*}{Eq.\ref{eq:lum_prob} ($t_b$=0.55, $t_d$=0.45)} & $5\times$\textbf{\textit{C\_100DW}} in Tab.\ref{tab:init_lut}                                                                                   & 33.906                                 & 0.9575                             & 16.999                           & 8.9071                             \\ \cline{2-2} \cline{5-9} 
			\multicolumn{1}{|c|}{}                                                                                                                         & \textbf{3C} &                                   &                                         & 4$\times$identity + 1$\times$all-1                                                                                     & 33.642                                 & 0.9582                             & 17.399                           & 8.9906                             \\ \cline{1-2} \hline 
		\end{tabular}
		\label{tab:abl}
	\end{table*}
	
	As in Tab.\ref{tab:abl}, we first investigate the impact of 2 core ingredients of our method: non-uniform \textit{precision} redistribution (Eq.\ref{eq:vertices}) and contribution map (Eq.\ref{eq:lum_prob}) guiding the redistribution of \textit{content}.
	
	\subsubsection{On non-uniform packing}
	\label{abl1}
\begin{figure}[h]
		\centering
		\includegraphics[width=\linewidth]{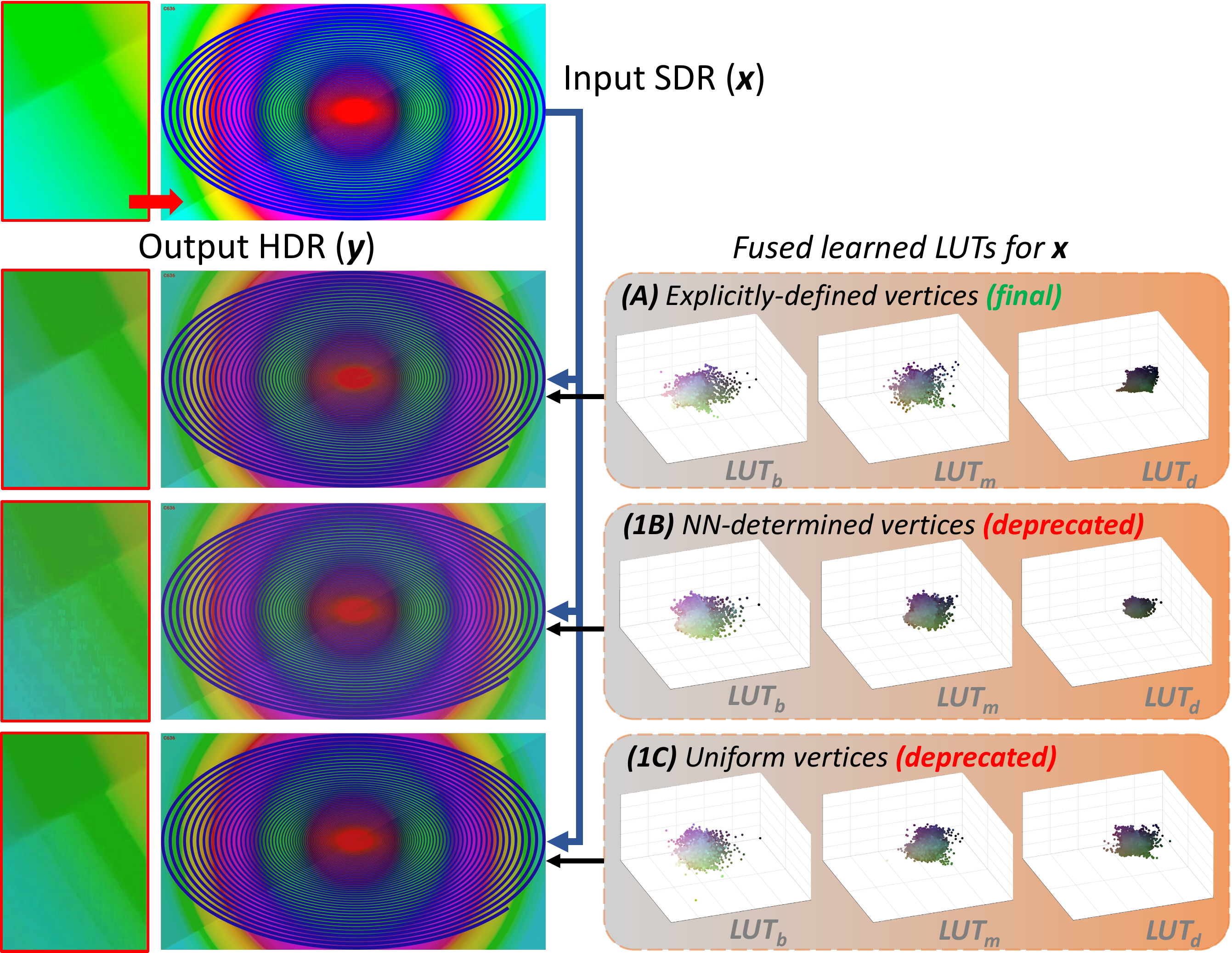}
		\caption{The impact of \textit{precision} redistribution ($v_l$, Eq.\ref{eq:vertices}): it will affect less the redistribution of $LUT_l$'s \textit{content}, yet more on objective metrics in Tab.\ref{tab:abl}.}
		\label{fig:non_uniform}
	\end{figure}

	The first alternative `\textbf{1B}' keeps vertices non-uniform, yet changes the source of non-uniformity from Eq.\ref{eq:vertices} to 3 neural networks respectively assigned to 3 branches (same structure and initialization as \cite{22-AdaInt}).
	As seen, metrics drop more than `\textbf{1C}' (Tab.\ref{tab:abl}) and artifact appears (Fig.\ref{fig:non_uniform} red box).
	The reason is that network-determined non-uniformity is hard to follow the assumed emphasis without a proper training constrain, and will thus conflict with the intention of Eq.\ref{eq:lum_prob}, which will be demonstrated in supplementary material.
	Then, we change vertices back to uniform (\textbf{1C}).
	In this case, Fig.\ref{fig:non_uniform} appears a similar visual, yet metrics in Tab.\ref{tab:abl} drop slightly.
	In brief, by showing the case of uniform (\textbf{1C}) and conflicting-non-uniform (\textbf{1B}), we prove the effectiveness of non-uniform \textit{precision} redistribution of Eq.\ref{eq:vertices}. 
	
	\subsubsection{On contribution map}
	\label{abl2}
	Here, 4 alternatives are:
	(1) Changing threshold $t_b$ and $t_d$ in Eq.\ref{eq:lum_prob} \textit{i.e} each branch's contribution: increasing the contribution of $\mathbf{y}_m$ (reducing that of $\mathbf{y}_b$ and $\mathbf{y}_d$) slightly (\textbf{2B}) and significantly (\textbf{2C}).
	(2) Not assigning each $\mathbf{y}_l$ with distinct contribution (\textbf{2D}), this equal to simply stacking the number of basic LUTs.
	(3) Using soft-curves to determine contribution map (\textbf{2E}):
	\begin{equation}
		\label{eq:lum_prob_soft}
		\begin{array}{l} 
			\left\{\begin{array}{l}
				p_b(\mathbf{x})=1-\frac{log(1+\mu(1-\mathbf{x}))}{log(1+\mu)} \\ 
				p_d(\mathbf{x})=1-\frac{log(1+\mu\mathbf{x})}{log(1+\mu)} \\ 
				p_m(\mathbf{x}) = 1-p_b(\mathbf{x})-p_d(\mathbf{x}) 
			\end{array}\right. \ ,\mathbf{x}\in[0,1],\ \mu=5000
		\end{array} 
	\end{equation}
	where $p_b$+$p_m$+$p_d$=1 $\forall \ \mathbf{x}\in[0,1]$, and (4) Hard segmentation where specific pixel $\mathbf{y}_i$ comes from the result of only 1 $LUT_l$/$\mathbf{y}_l$ (\textbf{2F}).
	
	\begin{figure}[h]
		\centering
		\includegraphics[width=\linewidth]{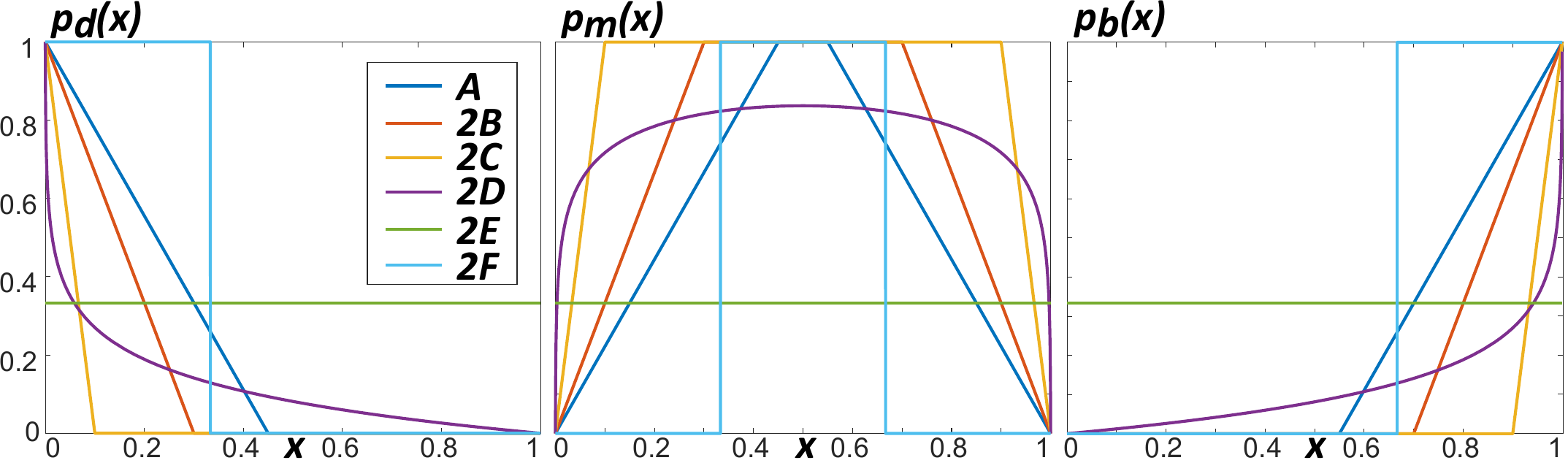}
		\caption{Luma probability functions $p_l$ (to generate contribution map from SDR $\mathbf{x}$) used in ablation study, \textbf{A} is original.}
		\label{fig:abl_lum_prob}
	\end{figure}
	
\begin{figure}[h]
		\centering
		\includegraphics[width=\linewidth]{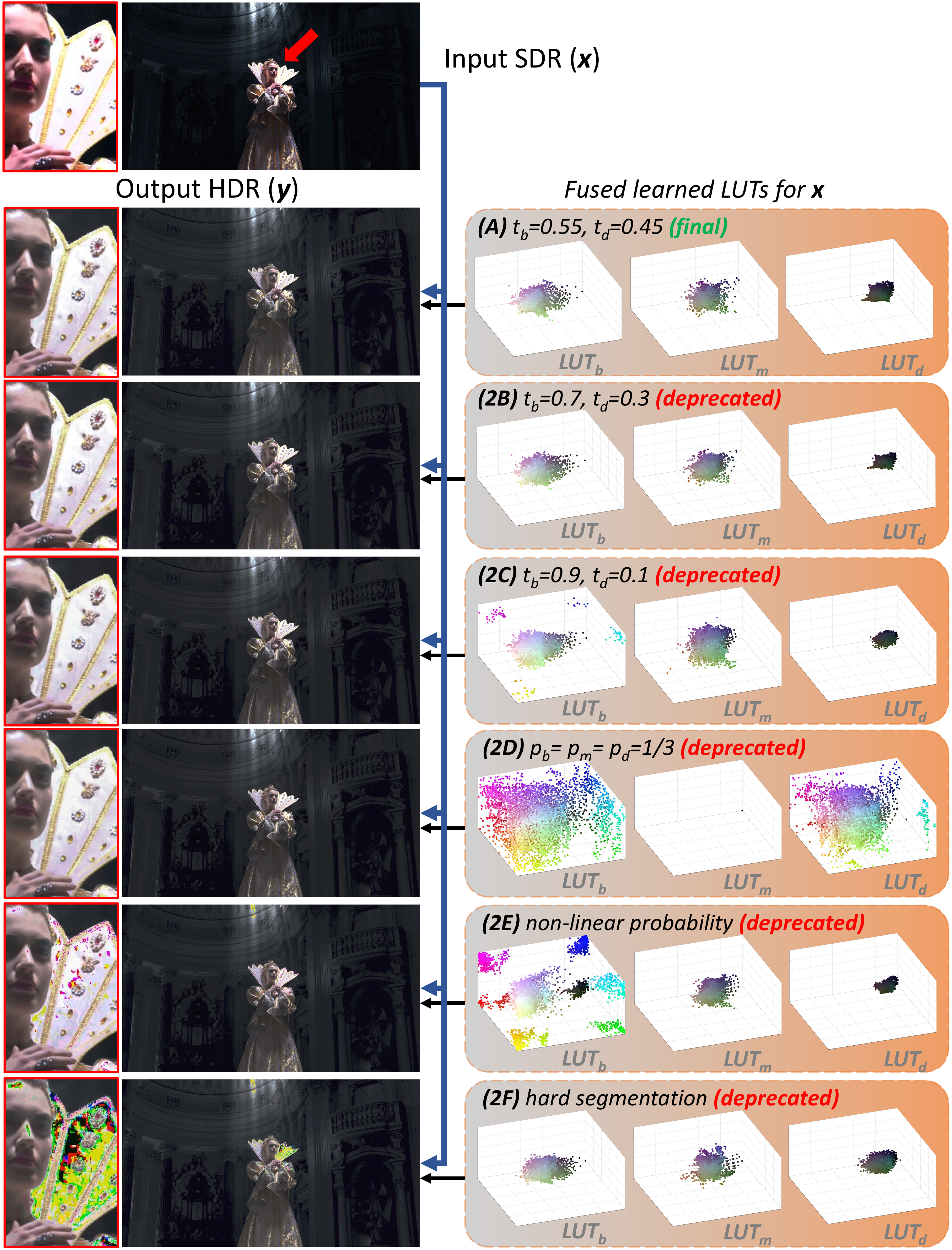}
		\caption{The effect of contribution redistribution ($p_l$, Eq.\ref{eq:lum_prob}): it will mostly change $LUT_l$'s learned \textit{content} thus method will map some SDR pixels to undesired color.}
		\label{fig:abl_lumprob}
	\end{figure}

	All alternative luma probability functions $p_l(\mathbf{x})$ are illustrated in Fig.\ref{fig:abl_lum_prob}.
	As in Fig.\ref{fig:abl_lumprob}, the worst case is \textbf{2D} whose $LUT_m$ is even trained to all-0-mapping (disabled).
	Eq.\ref{eq:lum_prob_soft} (\textbf{2E}) produces artifact since \textit{e.g.} $p_d$ there is also responsible for $\mathbf{y}$'s higher range which it shouldn't be.
	Also, the primitive way to combine $\mathbf{y}_b$, $\mathbf{y}_m$ and $\mathbf{y}_d$ to $\mathbf{y}$ (\textbf{2F}) affects less on the distribution of trained LUT content, yet produces more artifact (Fig.\ref{fig:abl_lumprob} red box).
	Lastly, keeping the shape of original probability curve Eq.\ref{eq:lum_prob} unchanged, increasing the contribution of $LUT_m$ slightly (\textbf{2B}) and more significantly (\textbf{2C}) will both affect less on \textit{content} distribution, yet Tab.\ref{tab:abl} indicate that the contribution of each $LUT_l$ should be in a reasonable (\textbf{A}\&\textbf{2B}) range.
	
	\subsubsection{On LUT initialization}
	\label{abl3}
\begin{figure*}[h]
		\centering
		\includegraphics[width=\linewidth]{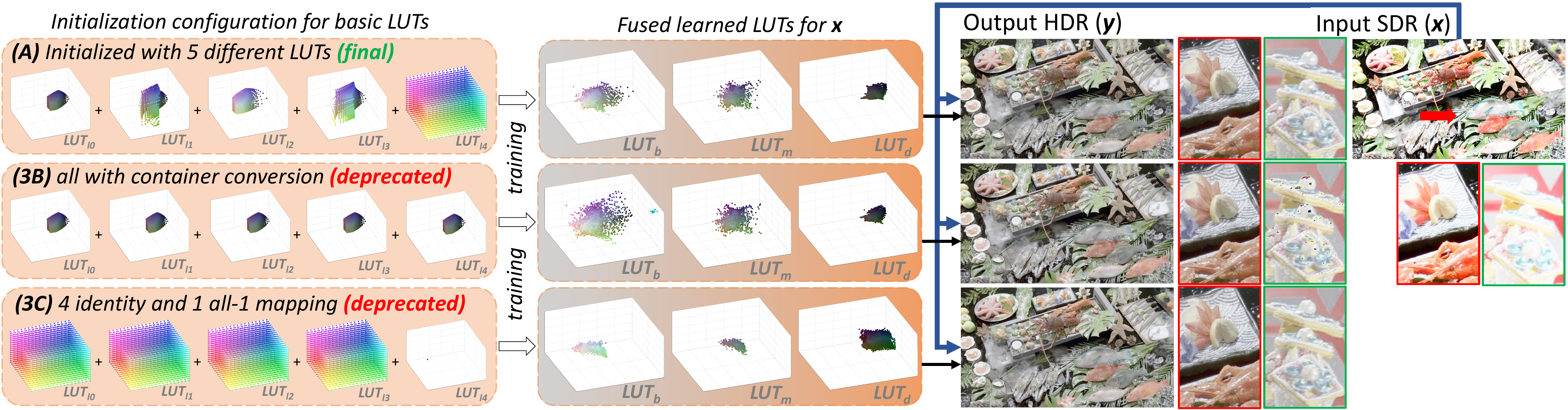}
		\caption{The role of basic LUTs initialization: it will mostly the learn LUT \textit{content} redistribution and produce color artifact if some parted of LUT is learned with undeserved mapping. Note that $LUT_{bi}$, $LUT_{mi}$ and $LUT_{di}$ ($i\in\{0,1,2,3,4\}$) are same initialized.}
		\label{fig:lut_init}
	\end{figure*}

	We then investigate LUT initialization whose particularity is brought by HDR/WCG, and is thus rarely discussed in SDR-oriented AI-LUT (Tab.\ref{tab:3d_luts}).
	First, in left Fig.\ref{fig:lut_init}, `\textbf{3B}' initializes basic LUTs with 5 container conversion (\textbf{\textit{C\_100DW}} in Tab.\ref{tab:init_lut}).
	It considers the HDR/WCG-SDR discrepancy, but is less diverse.
	Then, `\textbf{3C}' follows the same configuration as previous AI-LUT, and neglects the HDR/WCG particularity.
	
	From Fig.\ref{fig:lut_init} we know that `\textbf{2B}' affects less on the redistribution of LUT \textit{content} and more on visual artifact (green box), while effect of `\textbf{2C}' is more significant on \textit{content} redistribution and less on artifact.
	Also, from Tab.\ref{tab:abl} we know that `\textbf{2C}' drops more than `\textbf{2B}', this indicates that the amount of consideration on HDR/WCG discrepancy is positively correlated with method's performance.

	% ---------------------------------------------------------------
	\subsection{Objective Experiments}
	\label{sec:sota}
	
	\begin{table*}[t]
		\setlength{\abovecaptionskip}{0cm}
		\setlength{\belowcaptionskip}{0cm}
		\caption{The computation overhead of 5 AI-ITM, 3 AI-LUT (both LUT size $N$=33) and 3 AI-retouching methods.
			As seen, our method best meet the efficiency requirement of user-end display-oriented application.
		}
		\centering
		\small
		\begin{tabular}{|ccc|c|rr|rr|c|r|c|}
			\hline
			\multicolumn{3}{|c|}{\multirow{2}{*}{\textbf{Method}}}                                                                                                  & \multirow{2}{*}{\begin{tabular}[c]{@{}c@{}}Global\\ op. only\end{tabular}} & \multicolumn{2}{c|}{UHD resolution}                      & \multicolumn{2}{c|}{HD resolution}                       & \multicolumn{1}{c|}{\multirow{2}{*}{\begin{tabular}[c]{@{}c@{}}I2I\\ network\end{tabular}}}    & \multicolumn{1}{c|}{\multirow{2}{*}{\begin{tabular}[c]{@{}c@{}}\#net.\\ param.\end{tabular}}} & \multicolumn{1}{c|}{\multirow{2}{*}{\begin{tabular}[c]{@{}c@{}}\#LUT\\ element\end{tabular}}} \\ \cline{5-8}
			\multicolumn{3}{|c|}{}                                                                                                                                  &                                                                            & \multicolumn{1}{c|}{runtime} & \multicolumn{1}{c|}{GRAM} & \multicolumn{1}{c|}{runtime} & \multicolumn{1}{c|}{GRAM} & \multicolumn{1}{c|}{}                         &                                  &                                         \\ \hline
			\multicolumn{1}{|c|}{\multirow{5}{*}{AI-ITM}}                                                   & \multicolumn{1}{c|}{\textbf{Deep SR-ITM}\scriptsize{\cite{Kim19}}} & \scriptsize{ICCV'19}   & $\times$                                                                          & \multicolumn{2}{c|}{\scriptsize{N/A (GRAM OOM)}}                      & \multicolumn{1}{r|}{1.305s}  & 8099MB  & $\checkmark$                  & 2634k                                         & -                                                                         \\ \cline{2-11} 
			\multicolumn{1}{|c|}{}                                                                          & \multicolumn{1}{c|}{\textbf{SR-ITM-GAN}\scriptsize{\cite{Zeng20}}}  & \scriptsize{Access'20} & $\times$                                                                          & \multicolumn{1}{r|}{2.694s}  & 10539MB                   & \multicolumn{1}{r|}{0.608s}  & 5759MB  & $\checkmark$                  & 515k                                          & -                                                                         \\ \cline{2-11} 
			\multicolumn{1}{|c|}{}                                                                          & \multicolumn{1}{c|}{\textbf{HDRTVNet}\scriptsize{\cite{Chen211} \textbf{(ACGM+LE)}}}  & \scriptsize{ICCV'21}   & $\times$                                                                          & \multicolumn{1}{r|}{2.235s}  & 10165MB                    & \multicolumn{1}{r|}{0.559s}  & 5053MB  & $\checkmark$                  & 1404k                                         & -                                                                         \\ \cline{2-11} 
			\multicolumn{1}{|c|}{}                                                                          & \multicolumn{1}{c|}{\textbf{FMNet}\scriptsize{\cite{Xu222}}}       & \scriptsize{MM'22}     & $\times$                                                                          & \multicolumn{1}{r|}{2.635s}  & 8715MB                    & \multicolumn{1}{r|}{0.697s}  & 5843MB  & $\checkmark$                  & 1302k                                         & -                                                                         \\ \cline{2-11} 
			\multicolumn{1}{|c|}{}                                                                          & \multicolumn{1}{c|}{\textbf{LSN}\scriptsize{\cite{Guo23}}}         & \scriptsize{CVPR'23}   & $\times$                                                                          & \multicolumn{1}{r|}{1.554s}  & 9639MB                    & \multicolumn{1}{r|}{0.406s}  & 3867MB  & $\checkmark$                  & 325k                                          & -                                                                         \\ \hline
			\multicolumn{1}{|c|}{\multirow{3}{*}{AI-LUT}}                                                   & \multicolumn{1}{c|}{\textbf{A3DLUT}\scriptsize{\cite{20-3DLUT}} \textbf{($N$=33)}}      & \scriptsize{TPAMI'20}  & $\checkmark$                                                                          & \multicolumn{1}{r|}{0.279s}  & 2483MB                    & \multicolumn{1}{r|}{\textbf{0.051}s} & \textbf{2067}MB  & $\times$  & 296k                                          & \multicolumn{1}{r|}{323k}                                                 \\ \cline{2-11} 
			\multicolumn{1}{|c|}{}                                                                          & \multicolumn{1}{c|}{\textbf{AdaInt}\scriptsize{\cite{22-AdaInt}} \textbf{($N$=33)}}      & \scriptsize{CVPR'22}   & $\checkmark$                                                                          & \multicolumn{1}{r|}{0.310s}  & 3015MB                    & \multicolumn{1}{r|}{0.060s}  & 2286MB  & $\times$                  & 375k                                          & \multicolumn{1}{r|}{323k}                                                 \\ \cline{2-11} 
			\multicolumn{1}{|c|}{}                                                                          & \multicolumn{1}{c|}{\textbf{CLUT-Net}\scriptsize{\cite{22-CLUT-Net}} \textbf{($N$=33)}}    & \scriptsize{MM'22}     & $\checkmark$                                                                          & \multicolumn{1}{r|}{0.907s}  & \textbf{2365}MB           & \multicolumn{1}{r|}{0.249s}  & 2079MB  & $\times$                  & 264k                                          & \multicolumn{1}{r|}{\textbf{28k}}                                         \\ \hline
			\multicolumn{1}{|c|}{\multirow{3}{*}{\begin{tabular}[c]{@{}c@{}}AI-\\ retouching\end{tabular}}} & \multicolumn{1}{c|}{\textbf{NeurSpline}\scriptsize{\cite{Glob-Bianco20}}}  & \scriptsize{TIP'20}    & $\checkmark$                                                                          & \multicolumn{1}{r|}{1.087s}  & 3951MB                    & \multicolumn{1}{r|}{0.296s}  & 2463MB  & $\times$                  & 4457k                                         & -                                                                         \\ \cline{2-11} 
			\multicolumn{1}{|c|}{}                                                                          & \multicolumn{1}{c|}{\textbf{DeepLPF}\scriptsize{\cite{DeepLPF}}}     & \scriptsize{CVPR'20}   & $\times$                                                                          & \multicolumn{1}{r|}{1.652s}  & 8038MB                    & \multicolumn{1}{r|}{0.401s}  & 3483MB  & $\times$                  & 1769k                                         & -                                                                         \\ \cline{2-11} 
			\multicolumn{1}{|c|}{}                                                                          & \multicolumn{1}{c|}{\textbf{CSRNet}\scriptsize{\cite{Glob-He20}}}      & \scriptsize{TPAMI'22}  & $\checkmark$                                                                          & \multicolumn{1}{r|}{0.958s}  & 10325MB                   & \multicolumn{1}{r|}{0.608s}  & 5759MB & $\checkmark$                   & \textbf{35k}                                  & -                                                                         \\ \hline
			\multicolumn{1}{|c|}{ours}                                                                      & \multicolumn{2}{c|}{\textbf{ITM-LUT} \textbf{\scriptsize{(3$\times$$N$=17)}}}                 & $\checkmark$                                                                          & \multicolumn{1}{r|}{\textbf{0.254}s}  & 2869MB           & \multicolumn{1}{r|}{0.063s}  & 2145MB & $\times$                   & 452k                                          & \multicolumn{1}{r|}{221k}                                                 \\ \hline
		\end{tabular}
		\label{tab:overhead}
	\end{table*}
	
	\begin{table*}[t]
		\setlength{\abovecaptionskip}{0cm}
		\setlength{\belowcaptionskip}{0cm}
		\caption{Re-trained results of both global and non-global, AI-ITM, AI-LUT and AI-retouching methods.
			As seen, our method can reach acceptable performance compared with others with sophisticated network.
		}
		\centering
		\small
		\begin{tabular}{|cc|rrrr|rrrr|rrrr|}
			\hline
			\multicolumn{2}{|c|}{\multirow{2}{*}{\textbf{Method}}}                                                                 & \multicolumn{4}{c|}{Conventional metrics (distance to GT)}                                                                              & \multicolumn{4}{c|}{Expansion of HDR/WCG volume (\%)}                                                              & \multicolumn{4}{c|}{Style/aesthetic statistics (\%)}                                                             \\ \cline{3-14} 
			\multicolumn{2}{|c|}{}                                                                                                 & \multicolumn{1}{c|}{PSNR(dB)} & \multicolumn{1}{c|}{SSIM}   & \multicolumn{1}{c|}{$\Delta E_{itp}$}      & \multicolumn{1}{c|}{VDP3} & \multicolumn{1}{c|}{FHLP}   & \multicolumn{1}{c|}{EHL}    & \multicolumn{1}{c|}{FWGP}   & \multicolumn{1}{c|}{EWG} & \multicolumn{1}{c|}{ALL}    & \multicolumn{1}{c|}{Contrast}  & \multicolumn{1}{c|}{ASL}    & \multicolumn{1}{c|}{CF} \\ \hline
			\multicolumn{1}{|c|}{\multirow{5}{*}{AI-ITM}}                                                   & \textbf{Deep SR-ITM} & \multicolumn{1}{r|}{32.804}   & \multicolumn{1}{r|}{0.9373} & \multicolumn{1}{r|}{23.570}  & 8.8861                    & \multicolumn{1}{r|}{3.5562} & \multicolumn{1}{r|}{0.1558} & \multicolumn{1}{r|}{0.5503} & 0.1782                   & \multicolumn{1}{r|}{20.521} & \multicolumn{1}{r|}{81.345} & \multicolumn{1}{r|}{9.1469} & 9.6395                  \\ \cline{2-14} 
			\multicolumn{1}{|c|}{}                                                                          & \textbf{SR-ITM-GAN}  & \multicolumn{1}{r|}{32.659}   & \multicolumn{1}{r|}{0.9536} & \multicolumn{1}{r|}{20.609}  & 9.0901                    & \multicolumn{1}{r|}{2.3079} & \multicolumn{1}{r|}{0.0768} & \multicolumn{1}{r|}{0.5287} & 0.0799                   & \multicolumn{1}{r|}{18.482} & \multicolumn{1}{r|}{80.792} & \multicolumn{1}{r|}{8.5387} & 9.1991                  \\ \cline{2-14} 
			\multicolumn{1}{|c|}{}                                                                          & \textbf{HDRTVNet}  & \multicolumn{1}{r|}{33.523}   & \multicolumn{1}{r|}{0.9562} & \multicolumn{1}{r|}{18.756}  & 9.0964                    & \multicolumn{1}{r|}{5.0759} & \multicolumn{1}{r|}{0.5142} & \multicolumn{1}{r|}{0.3354} & 0.0952                   & \multicolumn{1}{r|}{20.798} & \multicolumn{1}{r|}{82.661} & \multicolumn{1}{r|}{9.7860} & 9.9985                  \\ \cline{2-14} 
			\multicolumn{1}{|c|}{}                                                                          & \textbf{FMNet}       & \multicolumn{1}{r|}{32.938}   & \multicolumn{1}{r|}{0.9264} & \multicolumn{1}{r|}{24.405}  & 9.0911                    & \multicolumn{1}{r|}{3.8903} & \multicolumn{1}{r|}{0.2969} & \multicolumn{1}{r|}{0.9793} & 0.1674                   & \multicolumn{1}{r|}{20.275} & \multicolumn{1}{r|}{85.030} & \multicolumn{1}{r|}{10.099} & 9.6456                  \\ \cline{2-14} 
			\multicolumn{1}{|c|}{}                                                                          & \textbf{LSN}         & \multicolumn{1}{r|}{\textbf{35.617}}   & \multicolumn{1}{r|}{\textbf{0.9667}} & \multicolumn{1}{r|}{\textbf{14.666}}  & \textbf{9.3839}                    & \multicolumn{1}{r|}{3.7387} & \multicolumn{1}{r|}{0.3885} & \multicolumn{1}{r|}{2.4938} & 0.1826                   & \multicolumn{1}{r|}{19.929} & \multicolumn{1}{r|}{81.453} & \multicolumn{1}{r|}{9.7997} & 9.9632                  \\ \hline
			\multicolumn{1}{|c|}{\multirow{3}{*}{AI-LUT}}                                                   & \textbf{A3DLUT}      & \multicolumn{1}{r|}{33.540}   & \multicolumn{1}{r|}{0.9539} & \multicolumn{1}{r|}{17.610}  & 8.8209                    & \multicolumn{1}{r|}{4.1767} & \multicolumn{1}{r|}{0.4965} & \multicolumn{1}{r|}{0.2151} & 0.0538                   & \multicolumn{1}{r|}{20.293} & \multicolumn{1}{r|}{85.536} & \multicolumn{1}{r|}{9.3331} & 9.2878                  \\ \cline{2-14} 
			\multicolumn{1}{|c|}{}                                                                          & \textbf{AdaInt}      & \multicolumn{1}{r|}{31.586}   & \multicolumn{1}{r|}{0.9378} & \multicolumn{1}{r|}{22.257}  & 8.5561                    & \multicolumn{1}{r|}{3.7596} & \multicolumn{1}{r|}{0.4617} & \multicolumn{1}{r|}{0.0666} & 0.0008                   & \multicolumn{1}{r|}{19.864} & \multicolumn{1}{r|}{85.639} & \multicolumn{1}{r|}{9.2363} & 8.9695                  \\ \cline{2-14} 
			\multicolumn{1}{|c|}{}                                                                          & \textbf{CLUT-Net}    & \multicolumn{1}{r|}{33.511}   & \multicolumn{1}{r|}{0.9569} & \multicolumn{1}{r|}{19.363}  & 8.9299                    & \multicolumn{1}{r|}{3.8815} & \multicolumn{1}{r|}{0.4780} & \multicolumn{1}{r|}{0.9013} & 0.2726                   & \multicolumn{1}{r|}{20.157} & \multicolumn{1}{r|}{85.457} & \multicolumn{1}{r|}{9.6476} & 10.441                  \\ \hline
			\multicolumn{2}{|c|}{Fixed LUT (\textbf{\textit{DaVinci}}, $N$=33)}                                                            & \multicolumn{1}{r|}{30.042}   & \multicolumn{1}{r|}{0.9529} & \multicolumn{1}{r|}{23.996}  & 9.0981                    & \multicolumn{1}{r|}{4.4174} & \multicolumn{1}{r|}{0.5429} & \multicolumn{1}{r|}{0.1964} & 0.0007                   & \multicolumn{1}{r|}{21.476} & \multicolumn{1}{r|}{85.172} & \multicolumn{1}{r|}{9.5059} & 9.2358                  \\ \hline
			\multicolumn{1}{|c|}{\multirow{3}{*}{\begin{tabular}[c]{@{}c@{}}AI-\\ retouching\end{tabular}}} & \textbf{NeurSpline}  & \multicolumn{1}{r|}{28.618}   & \multicolumn{1}{r|}{0.8545} & \multicolumn{1}{r|}{46.762}  & 8.8730                    & \multicolumn{1}{r|}{4.8400} & \multicolumn{1}{r|}{1.3018} & \multicolumn{1}{r|}{3.7184} & 0.9940                   & \multicolumn{1}{r|}{20.894} & \multicolumn{1}{r|}{79.410} & \multicolumn{1}{r|}{12.157} & 13.906                  \\ \cline{2-14} 
			\multicolumn{1}{|c|}{}                                                                          & \textbf{DeepLPF}     & \multicolumn{1}{r|}{29.774}   & \multicolumn{1}{r|}{0.8818} & \multicolumn{1}{r|}{30.977}  & 8.8153                    & \multicolumn{1}{r|}{2.5981} & \multicolumn{1}{r|}{0.1581} & \multicolumn{1}{r|}{1.0589} & 0.0774                   & \multicolumn{1}{r|}{19.735} & \multicolumn{1}{r|}{80.675} & \multicolumn{1}{r|}{10.121} & 10.326                  \\ \cline{2-14} 
			\multicolumn{1}{|c|}{}                                                                          & \textbf{CSRNet}      & \multicolumn{1}{r|}{31.970}   & \multicolumn{1}{r|}{0.9283} & \multicolumn{1}{r|}{22.887}  & 9.1484                    & \multicolumn{1}{r|}{3.1287} & \multicolumn{1}{r|}{0.2348} & \multicolumn{1}{r|}{0.5777} & 0.0348                   & \multicolumn{1}{r|}{19.493} & \multicolumn{1}{r|}{81.438} & \multicolumn{1}{r|}{9.7322} & 9.8361                  \\ \hline
			\multicolumn{1}{|c|}{ours}                                                                      & \textbf{ITM-LUT}     & \multicolumn{1}{r|}{34.203}   & \multicolumn{1}{r|}{0.9593} & \multicolumn{1}{r|}{17.013}  & 9.0363                    & \multicolumn{1}{r|}{4.0031} & \multicolumn{1}{r|}{0.5089} & \multicolumn{1}{r|}{0.6183} & 0.1897                   & \multicolumn{1}{r|}{20.255} & \multicolumn{1}{r|}{85.729} & \multicolumn{1}{r|}{9.4065} & 9.6538                  \\ \hline
			\multicolumn{2}{|c|}{\textbf{GT} HDR/WCG}                                                                              & \multicolumn{1}{c|}{-}        & \multicolumn{1}{c|}{-}      & \multicolumn{1}{c|}{-}       & \multicolumn{1}{c|}{-}    & \multicolumn{1}{r|}{4.8668} & \multicolumn{1}{r|}{0.7405} & \multicolumn{1}{r|}{2.5877} & 0.3610                   & \multicolumn{1}{r|}{20.749} & \multicolumn{1}{r|}{83.688} & \multicolumn{1}{r|}{9.7584} & 10.286                  \\ \hline
		\end{tabular}
		\label{tab:result}
	\end{table*}
	
	\begin{figure*}[h]
		\centering
		\includegraphics[width=\linewidth]{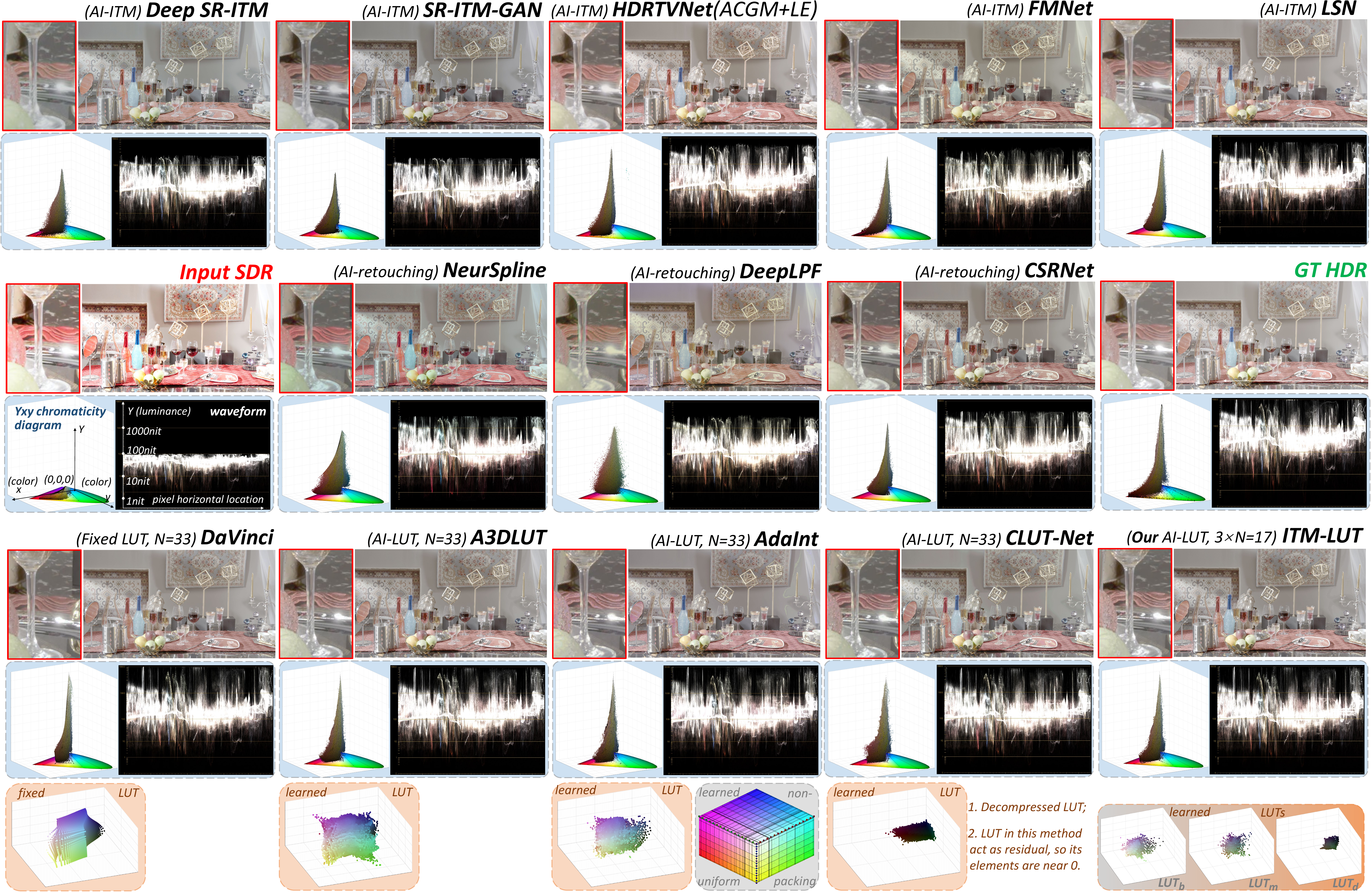}
		\caption{Visual comparison on highlight detail (red box), \textit{Yxy} chromaticity diagram and waveform (blue box) of both input SDR, result HDR/WCG and GT. Last row are (learned) LUTs of 5 LUT-based algorithms, as seen, they grab the SDR-HDR/WCG relation from training set in different ways. See supplementary material for more.}
		\label{fig:retrain}
	\end{figure*}
	
	\subsubsection{Computation overhead}
	For fairness, methods in Tab.\ref{tab:overhead} are all implemented by \textit{PyTorch}, re-trained by same training set, and executed on same \textit{Ubuntu} PC with i7-11700K CPU and A2000 GPU (12GB GRAM).
	Form Tab.\ref{tab:result} \textit{col.}5-8, single-frame runtime and GRAM consumption on UHD(3840$\times$2160) and HD(1920$\times$1080) resolution we know that: non-I2I(image-to-image, \textit{col.}9) and global (\textit{col.}4) AI methods generally execute faster.
	Among them, \textbf{our method} is the fastest under UHD \textit{i.e.} best for user-end scenario.
	Note that \textbf{CLUT-Net} is the slowest among AI-LUT, since its runtime is mostly spent decompressing its own compressed LUT representation.
	
	\subsubsection{Test set}
	We use 400 GT HDR/WCG from HDRTV4K\cite{Guo23} benchmark, and corresponding input SDR is `degraded' by \textit{DaVinci} as described earlier.
	Since SDR is less degraded, conventional metrics (PSNR, SSIM, $\Delta E_{itp}$ color difference\cite{BT2124} and VDP3\cite{VDP3}) will less manifest method's recover ability, but more on adaptability.
	As in Tab.\ref{tab:result}, \textbf{our method} got acceptable score compared with others with sophisticated network, which means global operation suffice display-end ITM.
	Also, \textbf{our method} surpassed other AI-LUT, which will verify our HDR/WCG-optimized 3-branch LUT design.
	Note that \textbf{AdaInt} underperforms, since its AI-determined non-uniform vertices conflict with HDR/WCG characteristics, as in bottom Fig.\ref{fig:retrain}.
	
	\subsubsection{HDR/WCG metrics}
	We use \textbf{f}raction/\textbf{e}xtent of \textbf{h}igh\textbf{l}ight/\textbf{w}ide-\textbf{g}amut (\textbf{p}ixels) (FHLP/EHL/FWGP/EWG) from \cite{Guo23} to measure result's HDR/WCG volume, and \textbf{a}vg. \textbf{l}uminance/\textbf{s}aturation \textbf{l}evel (ALL/ASL), contrast\cite{LuzardoITM} and colorfulness (CF)\cite{CF} for aesthetics.
	As seen, when methods are re-trained, they got similar HDR/WCG metrics. This indicate that the expansion of luminance and color is mostly done by the global $\mathbb{R}^3$$\rightarrow$$\mathbb{R}^3$ numerical mapping, rather the recovery ability provided by sophisticated network.
	Overall, as the fastest one, \textbf{our method} reached an acceptable expansion of HDR/WCG volume, and a good aesthetic consistency with GT.
	
	\subsubsection{Visuals}
	In Fig.\ref{fig:retrain}, all methods expand similar color/luminance (\textit{Yxy} diagram/waveform).
	An exception is AI-retouching \textbf{NeurSpline} and \textbf{DeepLPF} which have no cross-channel contamination, this proves the correctness of deprecating such global operations in ITM.
	While the rest methods share similar overall-look and highlight/dark detail when HDR/WCG is rendered as SDR.
	
	% ---------------------------------------------------------------
	\subsection{Subjective Experiments}
	\label{sec:subj}
	
	From Fig.\ref{fig:retrain}, we notice that SDR-rendered HDR/WCGs share little difference, yet, it will become significant when displayed on our Canon DP-V3120 HDR/WCG monitor.
	Therefore, we append a subjective study using this monitor and 10 footage from \cite{BT2245}: Each of 10 scenes is displayed as `GT-output1-GT-output2...' in random order, participants (9 total) are asked to rate from 0 to 100 based on how close current output is to GT, and for each rating they can optionally select attribution(s) from `overall brightness', `local contrast', `hue', `saturation' and `artifact'.
	
	The compared algorithms and GT-to-input degradation are same as objective study. Under such circumstance, \textbf{our method} outperforms all other fixed/AI-LUT, and underperforms only 2 of 5 AI-ITM algorithms (Due to space limit, please refer to supplementary material for result).
	This means our method will suffice user-end application whose SDR is with quality control (less degradation).
	
	% ===============================================================
	\section{Conclusion}
	\label{sec:conclu}
	This work combined LUT with AI for display-end ITM, and ameliorated it using the bright/middle/dark segmentation inspired by traditional ITM.
	Our method can: (1) learn a LUT in `\textit{bottom-up}' manner from any data, (2) execute fast and (3) alter with SDR statistics.
	Results have demonstrated our edge over AI-ITM, AI/fixed-LUT and AI-retouching methods.
	Currently the LUT size is 17, and our idea of redistributing \textit{precision} and \textit{content} can be applied to any other common size \textit{e.g.} 33 or 65, for an efficiency trade-off.
	
	Yet, there are still several aspects calling further study:
	(1) If there's a better redistribute of the non-uniformity of 3 smaller LUTs (Eq.\ref{eq:vertices})?
	(2) Is there a more reasonable contribution map (Eq.\ref{eq:lum_prob}) to cooperate with \textit{precision} redistribution (Eq.\ref{eq:vertices})?
	(3) Most importantly, the LUT \textit{content} is currently redistributed by the learning process, is there any way to explicitly define the \textit{content} redistributed, so our idea can be applied outside the learning/AI paradigm?
	Our method will be more practicable if these concerns are well addressed.

	%%
	%% The acknowledgments section is defined using the "acks" environment
	%% (and NOT an unnumbered section). This ensures the proper
	%% identification of the section in the article metadata, and the
	%% consistent spelling of the heading.
	\begin{acks}
		This work was supported in part by the National Natural Science Foundation of China under Grant 62101290.
	\end{acks}
	
	% \clearpage
	
	%%
	%% The next two lines define the bibliography style to be used, and
	%% the bibliography file.
	\bibliographystyle{ACM-Reference-Format}
	\bibliography{sample-bibliography2}

\end{document}